\documentstyle{article}
\begin{document}
\openup6pt

\title {Wormholes in vacuum Brans-Dicke theory} 
\author{Arunava Bhadra \thanks{Email address: aru\_bhadra@yahoo.com}  \\
Administrative Block \& IUCAA Reference Centre, 
University of North Bengal, Siliguri 734430 
INDIA \\
and \\
Kabita Sarkar  \\
Department of Mathematics, 
University of North Bengal, Siliguri 734430 
INDIA \\}
\date{}
\maketitle

\begin{abstract} 
It is shown that among the different classes of claimed static wormhole solutions of the vacuum Brans-Dicke theory only Brans Class I solution with coupling constant $\omega$ less than $-1.5$ (excluding the point $\omega =2$) gives rise to physically viable traversable wormhole geometry. Usability of this wormhole geometry for interstellar travel has been examined. 

\end{abstract}

PACS numbers:  04.20.Gz, 04.50.+h \\
\section{Introduction}
In recent years considerable interest has grown in the study of wormhole physics, either in general relativity or in alternative theories of gravitation, following the seminal work by Morris, Thorne [1] in which they introduced the concept of traversable wormholes and also obtained the properties that a space-time must have to hold up such geometry. Though the concept of wormhole came much earlier [2] as objects connecting different regions of spacetime but such wormholes were not traversable and thus were physically uninteresting. The idea of traversable wormholes opens up several possible interesting physical applications [3,4], for instance wormholes may be used as time machines [3]. A basic fact, however, is that for traversability it is essential to thread the wormhole throat with matter that violates the averaged null energy condition (ANEC). Most discussions of such exotic matter involve quantum field theory effects, such as the Casimir effect or Hawking evaporation. But, the quantum inequalities satisfied by the exotic matter fields tightly constrain the geometry of the wormhole by confining the exotic matter in a thin shell of size only slightly larger than the Planck length at the throat of the wormhole [5] thus essentially preventing the traversability. \\
The attempts to get around ANEC violation have led to increasing number of works in non-standard gravity theories such as the in the Brans-Dicke theory [6-9], $R+R^{2}$ theory [10], Einstein-Gauss-Bonnet [11], Einstein Cartan model [12], Kaluza-Klein [13] or in a Brane world scenario [14]. Though violation of ANEC is inevitable for traversability but some of these alternative theories allow one to use normal matter while relegating the exocity to non-standard fields. The study of wormhole geometry in the Brans-Dicke (BD) gravity theory [15], which describes gravitation through a spacetime metric ($g_{\mu\nu}$) and a massless scalar field ($\varphi$) that comes up naturally in most theoretical attempts at unifying gravity with other interactions or at quantizing gravity, receives special attention as the theory admits static wormholes both in vacuum [6-8] and with matter content that do not violate the ANEC by itself [9]. In this theory scalar field itself plays the role of the exotic matter and since it is a classical field, Roman-Ford restriction on the size of the traversable region is not applicable in this case. Although there are some ambiguities in the definition of energy density [16], it is generally considered that the energy density of scalar fields is not positive-definite in the BD theory [17]. But very recently it has been shown [18] that the Minkowski space is stable in this theory with respect to inhomogeneous scalar and tensor perturbations, at least at the linear order. This means negative energy associated with the scalar-tensor gravitational waves does not cause runway solutions at the classical level in the BD theory.  \\
The study of wormhole geometry in the BD theory has been initiated by Agnese and La Camera [6]. They have shown that the static spherically symmetric vacuum solution of the BD theory, which is often referred as Brans class I solution, gives rise to a two-way traversable wormhole for $\omega < -2$ where $\omega$ is the characteristic coupling constant of the theory. Since Birkhoff's theorem does not hold in the presence of a scalar field, several static solutions of the BD theory is possible even in spherically symmetric vacuum situation. Brans himself provided [19] four forms of static spherically symmetric vacuum solution of the BD theory (however as far as we know no other spherically symmetric solution that describes correctly the weak field observations is available in the literature). Among the all Brans classes of solutions Class I solution receives more attraction as it is the only one which is permitted for all values of $\omega$. The other three forms are valid only for $\omega <-3/2$ which implies non-positive contribution of matter to effective gravitational constant and thus a violation of the ANEC. Extending the work of Agnese and La Camera, Nandi {\it et al} showed [7] that several other Brans classes of solutions also support wormhole geometry. They further pointed out that  Brans class I solution admits wormhole geometry even when $\omega$ is positive. However, It has been shown recently [20] that only two of the Brans solutions are really independent and only class I solution represents exterior metric for a spherical gravitating object. Hence wormhole geometries corresponding to other classes of Brans solution are though mathematically viable but physically irrelevant; only wormhole geometry corresponding to Brans class I solution is physically meaningful. Recently He and Kim [8] have claimed for two {\it new} static vacuum wormhole solutions of the BD theory. But as already shown in [21, 20], these two classes of solutions are essentially limiting cases of the Brans class I solution. And moreover He-Kim classes of solutions do not satisfy all the standard weak-field observational results of gravitation. However, an important point of their analysis is that they have also considered the usability criteria [8] for effective  traversability. Such a study has not yet been considered in literatures in the context of the general and physical viable class of solution (Brans class I solution) of the BD theory.  \\  
For a wormhole to be traversable not just in principle but in practice it has to satisfy several usability criteria. For convenient travel tidal gravitational forces those a traveler feels must be bearably small, acceleration that the traveler experiences should not exceed much that of earth gravity as well as the time of journey to cross through the wormhole also must be finite and reasonable. In this article we would like to study wormhole geometry in vacuum BD theory considering both traversability and usability conditions as prescribed in [1]. We shall also impose the basic conditions of physical viability of the solution by demanding that the solution should represent external gravitational field for nonsingular spherical massive object and it must be consistent with the observational results. The present paper is organized as follows. After giving a short account of the BD theory and its static spherically symmetric solutions, physical viability of all Brans solutions will be discussed in section 2. In section 3 wormhole nature of the Brans solutions will be discussed by imposing traversability conditions. In section 4 we examine the usability of the wormhole under study for traveling to distant parts of the universe. Finally the results are discussed in section 5. \\

\section{Physically viable spherical symmetric vacuum solutions of the BD theory.}
In the BD theory, which accommodates both Mach's principle and Dirac's large number hypothesis, the scalar field acts as the source of the (local) gravitational coupling with $G \sim \varphi^{-1}$ and consequently the gravitational {\lq constant \rq}  is not in fact a constant but is determined by the total matter in the universe through an auxiliary scalar field equation. The scalar field couples to both matter and spacetime geometry and the strength of the coupling is represented by a single dimensionless constant $\omega$. The theory is consistent with the (local) observations only when $\omega$ is very large. A lower limit $\vert \omega \vert > 5 \times 10^4$ is imposed from the recent conjunction experiment with Cassini spacecraft [22]. Here it should be mentioned that in the limit $ \vert \omega \vert \rightarrow \infty$, the BD theory (and its dynamic generalization) reduces to GR unless the energy-momentum tensor is  traceless [23]. \\
In the Jordan conformal frame, the BD action takes the form (we use geometrized units such that $G=c=1$ and we follow the signature -,+,+,+)
\begin{equation}
{\cal A}= \frac{1}{16 \pi }\int d^{4}x \sqrt{-g}\left(\varphi R+\frac{\omega }{\varphi } 
g^{\mu\nu} \varphi_{,\mu} \varphi_{,\nu} + {\cal L}_{matter} \right)
\end{equation}
where \( {\cal L} _{matter} \) is the Lagrangian density of ordinary matter. Variation of (1) with 
respect to $ g^ {\mu \nu} $ and $\varphi$ gives, respectively, the field equations
\begin{equation} 
R_{\mu\nu} -\frac{1}{2}g_{\mu\nu}R= -\frac{8 \pi} {\varphi } T_{\mu \nu}
-\frac {\omega}{\varphi ^{2}}\left( \varphi_{,\mu} \varphi_{,\nu}- \frac{1}{2} g_{\mu\nu} 
\varphi^{,\sigma} \varphi_{,\sigma} \right) - 
\frac{1}{\varphi} \left( \nabla_{\mu}\nabla_{\nu}\varphi-g_{\mu \nu} \Box \varphi \right),  
\end{equation}
\begin{equation}
\Box \varphi = \frac {8\pi T}{(2\omega + 3)} 
\end{equation}
where $R$ is the Ricci scalar, and $T$=$T_{\mu}^{\mu}$ is the trace of the matter energy momentum tensor. \\
As stated earlier, though there are four Brans classes of static spherically symmetric solutions of the above theory when $T_{\mu \nu}=0$, only two of them are actually independent. The Brans class I solution (in isotropic coordinates) is given by 
\begin{equation}
ds^{2}= \left( \frac{1-B/\rho}{1+B/\rho} \right)^{\frac{2}{\lambda}} dt^{2} - \left( 1 + \frac{B}{\rho}\right) ^{4} \left( \frac{1-B/\rho}{1+B/\rho} \right)
^{\frac{2(\lambda -C -1)}{\lambda}} \left( d\rho^{2} +\rho^{2} d\theta ^{2} +\rho^{2} sin^{2} \theta d\phi ^{2} \right)
\end{equation}
\begin{equation}
\varphi = \varphi_{0} \left( \frac{1-B/\rho}{1+B/\rho} \right)^{\frac{C}{\lambda}}
\end{equation}
where $B, C, \lambda$ are arbitrary constants and the parameters are connected through the constraint 
\begin{equation}
\lambda^2 =  (C+1)^{2} - C(1-\frac{\omega C}{2})
\end{equation}
On the other hand the class IV solution reads
\begin{equation}
ds^{2}= e^{-2/(B\rho) } dt^{2} - e^{2(C+1)/(B\rho ) }\left( d\rho^{2} +\rho^{2} d\theta ^{2} +\rho^{2} sin^{2} \theta d\phi ^{2} \right)
\end{equation}
\begin{equation}
\varphi = \varphi_{0} e^{C/(B\rho)}
\end{equation}
with
\begin{equation}
C= \frac{-1 \pm \sqrt{-2\omega -3}}{\omega +2}
\end{equation}
Choice of imaginary $B$ and $\lambda$ in class I solution leads to the Brans class II solution [20]. A point to be noted is that under these choices the solution becomes regular at all points including the point $r=B$ and consequently the (class II) solution does not possess any horizon. Brans class III and class IV solutions are also not different [20,24]; under a mere redefinition of the radial variable ($\rho \equiv 1/\rho$) one of them maps to another. Both class I and class IV solutions are compatible with all the standard (up to the first post-Newtonian order) experimental tests of gravity conducted till now. The class I solution, which is the best known spherical symmetric solution of the BD theory, (in the Einstein conformal frame the corresponding solution is the well known Buchdahl solution [25] which is also variously referred [26] to as JNW [27] or Wyman solution [28]) in general gives rise to naked singularity whereas class IV solution is supposed to give rise the so-called cold black hole [29]. \\
In the quest for the physical viability of the solution we shall examine whether they match with the interior solution of the theory due to any reasonable spherical distribution of matter [20]. In general relativity the metric tensor is the only gravitational field variable. Hence, it is sufficient and necessary to match the interior and exterior solutions for metric tensor only. In contrast BD theory has additional scalar field which contributes to the gravitational field as well. Therefore, in this theory not only matching for metric tensor is necessary but also for the additional scalar field [30]. This is because the scalar field in Brans-Dicke theory (and also for its dynamic generalization) represents strength of gravitational field and locally measurable value of gravitational constant $G$ is a function of background scalar field $\varphi$. Since at the boundary there can be only one measured value of $G$ the interior and exterior scalar field has to be same there. \\
Moreover, metric tensor depends on the scalar field. For example the relation $\left( ln \varphi \right)_{,i}=K \left( ln g_{oo}^{1/2}\right)_{,i}$ [31], where $K$ is a constant, holds for static spherically symmetric solutions of the Brans-Dicke theory. This relation was derived by taking energy-momentum tensor of matter to that of a perfect fluid and the relation is unique provided the space-time is asymptotically flat and $\frac{\varphi}{g_{oo}}$ tends uniformly to a limit at infinity and its second derivative exists everywhere [31]. So mismatching of background field at the boundary surface results mismatching of metric and hence of geometry. \\
It follows from Eq. (3) that to the leading order in $1/r$ the interior (in presence of matter) scalar field satisfies the equation
\begin{equation}
\nabla^{2} \stackrel{2}{\varphi} =  -\frac{8\pi}{2\omega+3}\stackrel{o}{T} 
\end{equation}
where we have expanded scalar field as $\varphi = \varphi_{o} + \stackrel{2}{\varphi}+ \stackrel{4}{\varphi}+ ...$,  $\stackrel{N}{\varphi}$ denotes the term in $\varphi$ of order $\frac{1}{r^{N/2}}$ (here we have followed the notation of [32]) and $\stackrel{o}{T}$  denotes the term in $T^{\sigma}_{\sigma}$ of order $\frac{1}{r^{3}}$ ($\stackrel{o}{T}^{oo}$ is the density of rest mass) [32]. Therefore to the leading order in $1/r$ (r being the radial variable) the scalar field at near the surface of the matter distribution has the following expression 
\begin{equation}
\varphi = \varphi_{o} - \left(\frac{2}{2\omega+3} \right) \phi
\end{equation}
where $\phi$ is the Newtonian potential defined through $\nabla^{2} \phi = 4\pi \stackrel{o}{T}$. At the surface and outside the source $\phi = -M/r$ where $M$ is the total mass (a positive definite quantity) of the source as viewed by a distant observer. Utilizing the relation $G=\frac{2\omega+4}{2\omega+3}\frac{1}{\varphi_{o}}$ [32] where $G$ is the gravitational constant that will be measured in a real experiment, we finally get the expression for scalar field to the leading order in $1/r$ near the surface of the matter distribution
\begin{equation}
\varphi = \varphi_{o} \left( 1+\frac{1}{\omega+2} \frac{M}{r} \right) 
\end{equation}
A physically viable external solution for scalar field must match smoothly with the above expression at the surface. \\
When matching to two different solutions on a common surface it is essential of choosing an appropriate coordinate system. The expression (12) is written in standard coordinates ($t, r, \theta$ and $\phi$). Hence the expression for scalar field in Brans class I or class IV solution needs to transform first in the standard coordinates for effective comparison. However, at the first order the standard and isotropic coordinates produce identical effects. Hence comparing the expression for scalar field of Brans class I solution with Eq. (12) we get 
\begin{equation}
C=-\frac{1}{\omega +2}
\end{equation}
Here we have employed $2B/\lambda=M$ which is obtained from the relation $\stackrel{2}{g}_{oo}= 2\phi$. Consequently the constraint condition (6) gives 
\begin{equation}
\lambda=\sqrt{ \frac{2\omega +3}{2\omega +4}}
\end{equation}
Thus Brans class I solution (including class II variant) may represent external gravitational field due to a reasonable matter field only when the parameters $C$ and $\lambda$ are given respectively by Eqs. (13) and (14). On the other hand scalar field of class IV solution can not be matched with interior solution at the surface as its 1st order term goes as $\omega^{-1/2}$ for large $\omega$ whereas boundary condition requires that it must be proportional to $\omega^{-1}$. Thus regularity conditions at the boundary suggest that class IV solution does not describe exterior gravitational field for a nonsingular spherical massive object. 

\section{Brans wormholes}
Agnese and La Camera [6] already studied wormhole nature of Brans class I solution exactly with the choice of C as given by Eq. (12). They use post-Newtonian values of the BD theory to fix the parameter C. However, they did not consider imaginary values of the solution parameters. On the other hand Nandi {\it et al} [7] studied all classes of solutions but they did not apply any restriction on the parameters except the constraint condition (6) and the requirement of obeying Newtonian result in the weak field limit such that $2B/\lambda=M$. Thus the relaxed condition on the parameter $\omega$ as obtained by them for holding up wormhole geometry is though mathematically correct but physically unacceptable. But more importantly none of these works consider the usability of the BD wormhole. Hence for completeness we re-investigate the problem below. \\

\subsection{Traversable BD wormhole}
To be a wormhole the solution must have a throat that connects two asymptotically flat regions of spacetime. To examine whether a throat exists or not it is of convenience to cast the metric into the Morris-Thorne canonical form [1]
\begin{equation}
ds^{2}=e^{2\chi (R)} dt^{2}-\left[1-\frac{b(R)}{R}\right]^{-1} dR^{2}-R^{2}\left( d \theta^{2}+sin^{2}\theta d \phi^{2} \right)
\end{equation}
where $R$ is the new radial coordinate, $\chi(R)$ is known as the redshift function and $b(R)$ is called as shape function. \\
The class I solution can be cast to the above form by defining a radial coordinate $R$ which is related with $r$ via the expression
\begin{equation}
R=r \left( 1+\frac{B}{r} \right)^{1+(C+1)/\lambda} \left( 1-\frac{B}{r} \right)^{1-(C+1)/\lambda}  
\end{equation}
The functions $\chi(R)$ and $b(R)$ are the given by [6, 7]
\begin{equation}
\chi(R)=\frac{1}{\lambda}ln \left[\frac{ 1-B/r(R)}{1+B/r(R)}\right]
\end{equation}
\begin{equation}
b(R)=R \left[1-\left(\frac{(r^{2}(R)+B^{2})-2r(R)B(C+1)/\lambda}{r^{2}(R)-B^{2}} \right)^{2} \right]
\end{equation}
The axially symmetric embedded surface $z=Z(R)$ shaping the wormhole's spatial geometry is obtained from
\begin{equation}
\frac{dz}{dR}= \pm \left[ \frac{R}{b(R)} -1 \right] ^{-1/2}
\end{equation}
By definition of wormhole at throat its embedded surface is vertical. Hence the expression for the throat of the BD wormhole, which occurs at $R=R_{o}$ such that $b(R_{o})=R_{o}$, is given in r-coordinate by  
\begin{equation}
r_{o}^{\pm}=\frac{B}{\lambda}\left[ C+1 \pm \sqrt{ (C+1)^{2} - \lambda^{2}}\right]
\end{equation}
The choice (13) leads the above equation as 
\begin{equation}
r_{o}^{\pm}=\frac{M}{2(\omega +2)}\left[ \omega+1 \pm \sqrt{ -(3\omega+4)/2} \right]
\end{equation}
The throat radius thus becomes real when $\omega < -4/3$. However, positivity of the throat radius requires $\omega$ to be less than $-3/2$. An interesting range is $-2 <\omega < -1.5$ for which $\lambda$ becomes imaginary. But as shown in [20]  imaginary $\lambda$ together with imaginary $B$ leads to Brans class II solution. This  range thus gives rise viable wormhole geometry. The redshift function has a singularity at $r=B$ which corresponds to the point $R=0$. Hence traversability requires $R_{o}>0$ or equivalently $r_{o}>B$. This condition is satisfied when $r_{o}^{+}$ is the throat radius. The class I solution (including the class II variant) thus represents traversable two-way wormhole with throat radius $r_{o}^{+}$ when $\omega < -3/2$ excluding the value $\omega=2$.  \\
The expressions for wormhole geometry corresponding to the class II solution can be easily obtained from those of class I solution by replacing $B$ and $\lambda$ by $iB$ and $-i\Lambda$ respectively. This gives 
\begin{equation}
\chi(R)=\frac{2}{\Lambda}tan^{-1}(B/r)
\end{equation}
\begin{equation}
b(R)=R \left[1-\left(\frac{r^{2}(R)-B^{2}+2r(R)B(C+1)/\Lambda}{r^{2}(R)+B^{2}} \right)^{2} \right]
\end{equation}
and
\begin{equation}
r_{o}^{\pm}=\frac{B}{\Lambda}\left[- (C+1) \pm \sqrt{ (C+1)^{2}+\Lambda^{2}}\right]
\end{equation}
These are the expressions derived in [7] (the expression for shape function in [7] contains a minor error which is probably due to printing mistake) starting from {\it ab initio}. Unlike class I solution here throat radius is always real for any choice of $\Lambda$ and $C$. The $\omega$ dependence of $C$ (13) leads to the following expression for throat radius
\begin{equation}
r_{o}^{\pm}=\frac{M}{2(\omega +2)}\left[(\omega+1)\mp \sqrt{ -(3\omega+4)/2}\right]
\end{equation}
(here Newtonian limit leads the choice $2B/\lambda=-M$). As in the case of class I solution here also the throat radius  becomes real when $\omega < -4/3$ and becomes also positive only when $\omega < -3/2$. However, the range $\omega <-2$ makes $\lambda$ imaginary which in turns (together with imaginary $B$) maps the solution to class I solution and hence no new wormhole geometry is available. So class II solution only supports wormhole geometry when $-2<\omega<-3/2$. \\

\subsection{He-Kim classes of solutions}
The claimed new class I solution in [8] is essentially a limiting case of the Brans class II solution that can be obtained with the choice $\Lambda \rightarrow \infty$ and $\frac{C}{\Lambda} \equiv C$ as already shown in [21] and the expressions for the shape and redshift functions and the radius of the throat thus follow from Eqs.(22) -(24). The expression for the redshift function as given by Eq.(17) of [8] contains a sign anomaly which in turns affected the expression (Eq.(18) of [8]) for the radius of the throat. The correct expressions for the redshift function and radius of the throat for the He-Kim class I solution would be
\begin{equation}
b(R)=R \left[1-\left(\frac{(r^{2}(R)-B^{2})+2r(R)BC}{r^{2}(R)+B^{2}} \right)^{2} \right]
\end{equation}
and
\begin{equation}
r_{o}^{\pm}=BC\left[- 1 \pm \left( 1+\frac{1}{C^{2}} \right)^{1/2} \right]
\end{equation}
But anyway these solutions do not represent exterior gravitational field due to any reasonable spherically symmetric matter distribution and hence not physically viable.

\section {Usability}
As mentioned already for practical traversability across a wormhole several conditions, as prescribed in [1], have to comply which imposes restrictions on the geometry of the wormhole. Here additional restriction comes from the fact that the theory is consistent with the (local) observations only when $\omega$ is very large. 

\subsection{Feasible throat radius}
The throat radius has to be reasonably large for passage through wormhole. Insisting that $R_{o}$ should be at least of the order of $1 \; m$ we get the following condition on $\omega$ from Eqs. (20) invoking Eqs. (13) and (14)  in the limit of large $\omega$ 
\begin{equation}
\omega \le 1.3 \times 10^{7} \left( \frac{M}{M_{\odot}} \right)^{2}
\end{equation} 
So feasible throat radius demands $\omega$ to be bounded from the upper and the upper limit depends on the mass of the wormhole. The present observational restriction on $\omega$ thus sets a lower limit on the mass of the traversable BD wormhole.   \\ 

\subsection{Tidal forces}
Major constraints on wormhole geometry come from the tidal forces those a traveler feels while traveling across the wormhole. The tidal acceleration between two extreme parts (e.g. head to feet) of the traveler's body is given by [1]
\begin{equation}
\Delta a^{\i}= - R^{i'}_{0' k' 0'}\xi^{k'}  
\end{equation}
where $\xi$ is the vector separation between two extreme parts of the traveler's body , and $R^{i'}_{0' k' 0'}\xi^{k'}$ (Latin indices represents spatial components) are the components of the Riemann curvature tensor in the traveler's frame (denoted as primed) which are connected with those of the static observer's frame as follows 
\begin{equation}
R^{1'}_{0' 1' 0'}=R^{1}_{0 1 0}  
\end{equation}
and
\begin{equation}
R^{2'}_{0' 2' 0'}= R^{3'}_{0' 3' 0'}=\gamma^{2} R^{2}_{0 2 0} + \gamma^{2} v^{2} R^{2}_{1 2 1}  
\end{equation}
Here $v(R)$ is the radial velocity of the traveler when he/she passes the radial point r, as measured by a static observer there, $\gamma$ is the usual Lorentz factor ($\gamma^{2} \equiv \frac{1}{1-v^{2}}$). For a convenient wormhole travel by  human beings ($\xi \sim 2 \; m$) tidal accelerations should not exceed much that of earth gravity ($g_{\oplus}$) i.e.,   
\begin{equation}
\vert R^{1'}_{0' 1' 0'} \vert \le \frac{1}{(10^{8} \; m)^{2} }  
\end{equation}
\begin{equation}
\vert R^{2'}_{0' 2' 0'} \vert \le \frac{1}{(10^{8} \; m)^{2} }  
\end{equation}
The constraint (32) arising from radial tidal acceleration restricts the wormhole geometry whereas the constraint (33) which is due to lateral tidal acceleration imposes restriction on the velocity of the traveler. In the static observer's frame the relevant non-vanishing components of the Riemann tensor for the Brans class I metric are given by   
\begin{equation}
R_{0101} = -\frac{4 B r^3 \big(\frac{r-B}{r+B}\big)^{\frac{2 (C+1)}{\lambda }} ( (B^{2} +r^{2}) \lambda -(C+2) B r   )}{(r-B)^{4} (r+B)^{4} \lambda ^{2}}
\end{equation}
\begin{equation}
R_{0202} = \frac{2 B r^3 \big(\frac{r-B}{r+B}\big)^{\frac{2 (C+1)}{\lambda }} ((B^{2} +r^{2}) \lambda -2 (C+1) B r   )}{(r-B)^{4} (r+B)^{4} \lambda ^{2}}
\end{equation}
and
\begin{equation}
R_{1212} = \frac{2 B r^3 \big(\frac{r-B}{r+B}\big)^{\frac{2 (C+1)}{\lambda }} ((C+1) (B^{2}+r^{2})  -2 B r \lambda ) }{(r-B)^{4} (r+B)^{4} \lambda }
\end{equation}
For physical viability the parameters $C$ and $\lambda$ are to be substituted by Eqs. (13) and (14). \\
It may appear from the expression of $R_{1' 0' 1' 0'}$ that due to the presence of $(r-B)^{4}$ in the denominator the tidal acceleration would become very large at throat but since observations already constrain $1/\omega$ to a very small value,  $R_{1' 0' 1' 0'}$ effectively tends (excluding the exact point $r=B$) to GR expression $-\frac{4Br^{3}}{(r+B)^{6}}$  which is finite. Thus the condition (32) leads to the constraint $M >10^{4} M_{\odot}$. Hence mass of the wormhole must be very large for effective traversability. \\   
On the other hand the condition (33) gives
\begin{equation}
\gamma^{2} \left[ \frac{2 B r^3 \big(\frac{r-B}{r+B}\big)^{\frac{2 (C+1)}{\lambda }} }{(r-B)^{4} (r+B)^{4} \lambda } \right] \left[ (B^{2} +r^{2}) (\lambda -v^{2} (C+1) ) -2 B r (C+1-v^{2}\lambda) \right]  \le  \frac{1}{(10^{8} \; m)^{2} }
\end{equation}
As before in the limit of large $\omega$ the left hand side of the above equation effectively reduces to $\frac{2Br^{3}}{(r+B)^{6}}$  which is finite and is interestingly independent of velocity of the traveler. Hence essentially there is no restriction on the velocity of the traveler from tidal accelerations. \\

\subsection{Other dynamical constraints}
The acceleration that a traveler would sense (note that traveler is not freely falling) in passing through a gravitational field is given by [1]
\begin{equation}
a=g_{oo}^{-1} \frac{d (\gamma g_{oo})}{dl}
\end{equation}
The demand that traveler should not feel acceleration greater than about 1 Earth gravity while traveling through the wormhole leads to the following condition for the BD wormhole
\begin{equation}
\frac{(r^{2}+B^{2})\lambda-2rB(C+1)}{\lambda(r^{2}-B^{2})} \left[ \gamma \frac{2B}{\lambda (r^{2}-B^{2})}-\gamma^{'} \right] \le 1.1 \times 10^{-16} \; m^{-1} 
\end{equation}
(here prime denotes the derivative with respect to $r$). The acceleration would be, thus, very large particularly near the throat and for large $\omega$ the above condition demands $\gamma (M/M_{\odot})^{-1} \le 2 \times 10^{-12}$ {\it i.e.}, mass of the wormhole needs to be abnormally large. \\ 
For practical traversability the total journey time to cross through the wormhole must be finite and reasonably small as measured by the traveler as well as persons waiting outside the wormhole. The journey time measured by people who live in the stations
\begin{equation}
\delta t=\int _{-l_{1}}^{l_{2}} \frac{dl}{v \; g_{oo}}
\end{equation}
where $l_{1}$ and $l_{2}$ coordinates of the space-stations. It is required that this time is of the order of $1$ year. Certainly this condition can not be met when the mass of the wormhole is abnormally large ($\sim 10^{12} M_{\odot}$). In that case space stations, where geometry of the spacetime must be nearly flat and acceleration of the gravity should be at maximum of the order of gravity, would be at large distances ($\sim 10^{7}$ pc) from the wormhole throat and even by moving with the speed of light the traveler would not be able to reach one station from another within a reasonable period of time. 

\section{Discussion}
Recently a measure of quantifying exotic matter needed for traversable wormhole geometry has been advanced [33] in the context of general relativity. It appears that no such quantifying measure is really needed in this case. As is evident from the foregoing analysis that the existence of traversable wormhole geometry for class I solution entirely depends on the mass and coupling constant $\omega$. As $1/\omega$ goes to zero the BD theory converges to the general relativity (leaving aside the case of traceless matter source), class I solution tends to Schwarzschild solution and two-way traversability disappears. But existence of scalar field (which is pre-assumed in the BD theory) with very small (bounded by Eq.(40)) negative $1/\omega$ is sufficient to give rise (in principle) traversable wormhole in the BD theory. \\
Traversability through wormhole requires violation of ANEC and hence exotic matter. Since all standard energy conditions can be violated easily at the classical level in a theory involving scalar field that coupled non minimally with the space-time [34] (in fact it has been shown recently [35] that a spontaneous violation of the energy conditions occurs for a wide class of scalar tensor theories as a consequence of spontaneous scalarization [30] thus providing a natural way by which normal matter could be transformed into exotic, in the interior of neutron stars.), BD theory admits traversable wormholes for particular choices of parameters. An important question is that whether BD vacuum wormholes are suited for interstellar travel or not. It has been shown above that the practical traversability restricts the mass of the wormhole and coupling parameter. The reasonable throat radius, which is required for passage through wormhole, demands $\omega$ to be bounded from upper. This upper limit depends on the mass of the wormhole and the present observational restriction on $\omega$ does not rule out a journey through BD wormholes. The tidal forces those a traveler would feel while traveling across the wormhole constrains the mass of the wormhole to a large value, at least $10^{4} \times M_{\odot}$. But most severe restriction on mass of the wormhole comes from the acceleration that a traveler would experience while approaching to the throat of the wormhole. This results a lower bound of the mass of the wormhole as $M > 5 \times 10^{12} M_{\odot}$. Consequently journey time to cross through the wormhole would be very large.  The present investigation, thus, shows that vacuum BD wormhole is though traversable in principle but not suitable for interstellar travel.\\

\noindent {\bf Acknowledgments}
Authors thank an anonymous referee for drawing attention to some useful references. \\


\begin{thebibliography}{99}
\bibitem{ab:1} M. S. Morris and K. S. Thorne {\it Am. J. Phys. } {\bf 56}, 395 (1988).
\bibitem{ab:2} J. A. Wheeler {\it Geometrodynamics } (Academic, New York, 1962) 
\bibitem{ab:3} M. S. Morris, K. S. Thorne and U. Yurtsever {\it Phys. Rev. Letts.} {\bf 61}, 1446 (1988); V. P. Frolov and I. D. Novikov, {\it Phys. Rev. D} {\bf 42}, 1057 (1990) 
\bibitem{ab:4} D. Hochberg and T. Kephart {\it Phys. Rev. Letts.} {\bf 70}, 2665 (1993); {\it Gen. Relativ. Gravit. } {\bf 26}, 219 (1994); V. P. Frolov and I. D. Novikov, {\it Phys. Rev. D} {\bf 48}, 1607 (1993); S. Coleman {\it Nucl. Phys. B} {\bf 310}, 643 (1988); S. Hawking, {\it Nucl. Phys. B} {\bf 335}, 155 (1990); J. G. Cramer, R. L. Forward, M. S. Morris, M. Visser, G. Benford, and G. A. Landis, {\it Phys. Rev. D} {\bf 51}, 3117 (1995) 
\bibitem{ab:5} L. H. Ford and T. A. Roman, {\it Phys. Rev. D} {\bf 51}, 4277 (1995); A. Borde, L. H. Ford and T. A. Roman, {\it Phys. Rev. D} {\bf 65}, 084002 (2002); L. H. Ford, A.D. Helfer and T. A. Roman, {\it Phys. Rev. D} {\bf 66}, 124012 (2002).
\bibitem{ab:6} A. G. Agnese and M. La Camera, {\it Phys.Rev.D}, {\bf 51}, 2011 (1995)
\bibitem{ab:7} K.K.Nandi, A.Islam, and J.Evans, {\it Phys.Rev. D}, {\bf 55}, 2497(1997)
\bibitem{ab:8}F. He and S-W. Kim, {\it Phys.Rev.D}, {\bf 65}, 084022 (2002)
\bibitem{ab:9} L. A. Anchordoqui, S. E. Perez Bergliaffa and D. F. Torres, {\it Phys. Rev. D}, {\bf 55}, 5226 (1997)  
\bibitem{ab:10} D. Hochberg, {\it Phys. Letts. B}, {\bf 251}, 349 (1990)  
\bibitem{ab:11}S. Kar, {\it Phys.Rev.D}, {\bf 53}, 722 (1996)
\bibitem{ab:12} L. A. Anchordoqui, {\it Mod.Phys.Lett.A}, {\bf 13},1095 (1998); L.C. Garcia de Andrade, {\it Mod. Phys. Lett. A},{\bf 15}, 1321 (2000)  
\bibitem{ab:13}Y.-G. Shen, H.-Y.Guo, Z.-Q. Tan, and H.-G. Ding, {\it Phys.Rev.D}, {\bf 44}, 1330 (1991)
\bibitem{ab:14} K. A. Bronnikov, and S.-W. Kim, {\it Phys. Rev. D}, {\bf 67}, 064027 (2003); L. A. Anchordoqui, and S. E. Perez Bergliaffa, {\it Phys. Rev. D}, {\bf 62}, 067502 (2000)  
\bibitem{ab:15} C.H. Brans and R.H. Dicke, {\it Phys. Rev. } {\bf 124}, 925 (1961).
\bibitem{ab:16}S. Bellucci, and V. Faraoni, {\it Nucl. Phys. B}, {\bf 640}, 453 (2002); D. I. Santiago, and A. S. Silbergleit, {\it Gen. Rel. Grav. } {\bf 32}, 565 (2001); D. F. Torres, {\it Phys. Rev. D} {\bf 66}, 043522 (2002)
\bibitem{ab:17}G. Magnano, and L. M. Sokolowski, {\it Phys. Rev. } D {\bf 50}, 5039 (1994); V. Faraoni, E. Gunzig, and P. Nardone, P. {\it Fundamentals of Cosmic Physics }, {\bf 20}, 121 (1998).
\bibitem{ab:18}V. Faraoni, {\it Phys.Rev.D}, {\bf 70}, 081501 (R) (2004)
\bibitem{ab:19}C.H. Brans , {\it Phys. Rev. }{\bf 125},2194 (1962).
\bibitem{ab:20}A. Bhadra and K. sarkar, {\it Gen. Rel. Grav.}, {\bf 37}, 2189 (2005) [gr-qc/0505141]
\bibitem{ab: 21} A.Bhadra, I. Simaciu, Y.-Z. Zhang and K.K. Nandi, {\it Phys.Rev.D}, {\bf 71}, 128501 (2005) [gr-qc/0406014]
\bibitem{ab: 22} B. Bertotti {\it et al}, {\it Nature } {\bf 425}, 374 (2003)
\bibitem{ab:23} A.Bhadra and K.K.Nandi, {\it Phys. Rev. D} {\bf 64}, 087501 (2001); C. Romero and A.Barros,  {\it Phys.Lett.} {\bf A173}, 243 (1993);  N.Banerjee and S.Sen, {\it Phys.Rev. D} {\bf 56},1334 (1997); V. Faraoni,{\it Phys.Rev.D} {\bf 59}, 084021 (1999); A. Bhadra, preprint gr-qc/0204014.
\bibitem{ab: 24} A.Bhadra, K.K.Nandi, {\it Mod. Phys. Letts.} {\bf 16} 2079 (2001).
\bibitem{ab: 25} H.A.Buchdahl, {\it Phys.Rev.},{\bf 115},1325 (1959)
\bibitem{ab: 26}A.Bhadra, K.K.Nandi, {\it Int. J. Mod. Phys. A} {\bf 16} 4543 (2001).
\bibitem{ab:27} A.I.Janis, E.T.Newman and J.Winnicour, {\it Phys. Rev. Lett.}, {\bf 20}, 878(1968).
\bibitem{ab:28} M.Wyman, {\it Phys.Rev. D}, {\bf 24}, 839(1981).
\bibitem{ab:29} K. K. Nandi, T. B. Nayak, A. Bhadra and P.M. Alsing, {\it Int. J. Mod. Phys. D}, {\bf 10}, 529 (2001).
\bibitem{ab:30}T. Damour and G. Esposito-Farese, {\it Phys. Rev. Letts.} {\bf 70}, 2220 (1993)
\bibitem{ab:31}W. F. Bruckman and E. Kazes, {\it Phys. Rev. D} {\bf 16}, 269 (1977)
\bibitem{ab:32} S.Weinberg, {\it Gravitation and Cosmology (Wiley, New York,1972)}
\bibitem{ab: 33} M. Visser, S. Kar and N. Dadhich, {\it Phys. Rev. Letts.} {\bf 90}, 201102 (2003)
\bibitem{ab: 34} C. Bercelo, and M. Visser, {\it Class. Quant. Grav. } {\bf 17}, 3843 (2000)
\bibitem{ab: 35}A. W. Whinnett, and D. F. Torres, {\it Astrophys. J.} {\bf 603}, L133 (2004) 

\end{thebibliography}
\end{document}